\documentclass[conference]{IEEEtran}

\usepackage[dvips]{graphicx}
\usepackage{amssymb,amsmath}
\usepackage{algorithm}
\usepackage{algorithmic}
\usepackage{cite}
\usepackage{hyperref}
\usepackage{flushend}
\usepackage{todonotes}

\begin{document}
\title{A Deep Learning Approach for Low-Latency Packet Loss Concealment of Audio Signals in Networked Music Performance Applications}
\date{}%date stay empty
%
%\author{
%\IEEEauthorblockN{Prateek Verma, Chris Chafe}
%\IEEEauthorblockA{Stanford University\\Stanford, California\\\{prateekv,\:cc\}@ccrma.Stanford.edu\\}
%\and
%\IEEEauthorblockN{Alessandro Ilic Mezza}
%\IEEEauthorblockA{Politecnico di Milano\\Milan, Italy \\alessandroilic.mezza@polimi.it\\}
%\and
%\IEEEauthorblockN{Cristina Rottondi}
%\IEEEauthorblockA{Politecnico di Torino\\Turin, Italy\\cristina.rottondi@polito.it\\}}
%
\author{
\IEEEauthorblockN{Prateek Verma$^\star$, Alessandro Ilic Mezza$^\dagger$, Chris Chafe$^\star$, Cristina Rottondi$^\ddagger$}
\IEEEauthorblockA{$^\star$\:Stanford University, Stanford, California\\$^\dagger$\:Politecnico di Milano, Milan, Italy\\$^\ddagger$\:Politecnico di Torino, Turin, Italy\\\{prateekv,\:cc\}@ccrma.Stanford.edu, alessandroilic.mezza@polimi.it, cristina.rottondi@polito.it}\\}
\maketitle

\begin{abstract}
Networked Music Performance (NMP) is envisioned as a potential game changer among Internet applications: it aims at revolutionizing the traditional concept of musical interaction by enabling remote musicians to interact and perform together through a telecommunication network.
Ensuring realistic conditions for music performance, however, constitutes a significant engineering challenge due to extremely strict requirements in terms of audio quality and, most importantly, network delay. To minimize the end-to-end delay experienced by the musicians, typical implementations of NMP applications use uncompressed, bidirectional audio streams and leverage UDP as transport protocol. Being  connectionless and unreliable, audio packets transmitted via UDP which become lost in transit are not retransmitted and thus cause glitches in the receiver audio playout. 
This article describes a technique for predicting lost packet content in real-time using a deep learning approach.
%The performance of the system is compared to that achieved by state-of-the-art autoregressive models.
The ability of concealing errors in real time can help mitigate audio impairments caused by packet losses, thus improving the quality of audio playout in real-world scenarios.
\end{abstract}

\section{Introduction}
\label{sec:intro}

Performing music together at distances across the Internet is a practice that makes use of low-latency uncompressed audio streaming technologies. According to several studies \cite{carot2009fundamentals,rottondi2015feature}, the delay tolerance threshold is estimated to be 20--30~ms, corresponding to a distance of approximately 8--9~m (considering the speed of sound propagation in air), a physical spacing which can be considered as a maximum separation ensuring the maintenance of a common tempo without a conductor. From the networking point of view, very strict requirements must therefore be satisfied to keep the one-way end-to-end transmission delay below a few tens of milliseconds.

A common solution sends UDP audio packets directly between a pair of network hosts (full-duplex, peer-to-peer), thus avoiding the processing time required by compression codecs and delays introduced by the packet re-transmission mechanisms supported by TCP, while sacrificing data transfer reliability. Despite having no guarantees that the audio exchange will be error-free, good networks and appropriate tuning of audio and network parameters can still result in connections which are quite satisfactory for music performance, and even for concert presentations. Experiences with distributed ensembles \cite{chafe2018frontiers} have included multi-channel audio and multi-site collaborations (greater than stereo and greater than two sites).  

In order to ensure optimal conditions for ensemble synchronization, a strict deadline is imposed on the arrival of each audio packet at the receiving site. The amount of time between when the packet was generated on the sender side and this playback deadline constitutes part of the time delay or lag  experienced by the collaborating musicians. Added to this is any propagation delay in air such as between instrument or voice and microphone, and loudspeaker to ear, as well as time spent in signal conversion between analog and digital. Packet streams transmitted across a Wide Area Network (WAN) require buffering at the receiver to account for jitter (i.e., variations of packet inter-arrival times). Critically, the buffer queue must be kept as low as possible since longer \lq\lq cushions'' (longer buffer queues) create longer transmission delays. A lost packet causes a gap in the playback audio, and an audio packet that arrives so late that the buffer queue has been exhausted is effectively a lost packet. Since the audio signal is not encoded prior to transmission, the receiving side cannot rely on error correction mechanisms implemented in state-of-the-art audio codecs. Therefore, alternative recovery algorithms must be devised to cope with packet losses, in order to mitigate the resulting audio glitches. Such solutions must operate in real time and without introducing additional processing delays that would worsen the end-to-end latency. 

In the reminder of this paper, we will focus on jacktrip \cite{caceres2010jacktrip}, which is an open-source project used for concert and online jamming. The jacktrip application belongs to a class of uncompressed audio streaming applications including Soundjack \cite{carot2008distributed}, LOLA \cite{drioli2013networked} and UltraGrid \cite{holub2012ultragrid}. If a packet is lost in a jacktrip connection, its built-in Packet Loss Concealment (PLC) algorithm simply repeats the last good packet that was received. Alternatively, its PLC mode can be set to mute the audio until the next good packet is able to be played out. In most cases, these methods of coping with signal errors are inappropriate and some degree of distortion is audible. 

 What to do to repair the gaps from lost packets is the subject of this paper, which proposes a method to predict the missing audio data in real-time by leveraging a deep learning framework. The Machine Learning (ML) algorithm which will be described is trained with audio samples extracted from a database of music which resembles the music that will be transmitted. The effectiveness of the proposed solution is evaluated by artificially introducing gaps in a test audio signal, inserting intentional errors which mimic in a controlled way the packet losses due to transmission through a WAN. The test algorithm predicts the missing audio and its performance is compared to a state-of-the-art AutoRegressive (AR) model used for PLC. Results show that, for a range of configuration parameters, the ML algorithm outperforms the AR approach in repairing audio signal gaps. 
 
 The remainder of this paper is organized as follows: after briefly reviewing the related literature in Section \ref{sec:related}, we provide some background notions on AR models in Section \ref{sec:background} and then describe the proposed ML framework in Section \ref{sec:framework}. A performance assessment is provided in Section \ref{sec:results}.

\section{Related Work}
\label{sec:related}
NMP over packet switched internets was first experimented with in the 90's and has gathered increasing attention in the scientific community in the last two decades. The widespread diffusion of internet services and their continuous performance improvements in terms of latency and capacity have fostered a music-ensembles-at-a-distance performance practice which is particularly timely at the time of this writing while the world copes with the SARS-CoV-2 pandemic and social distancing is required. For a thorough overview on NMP technologies, the reader is referred to \cite{7769205}.
Due to the best-effort paradigm of the IP protocol and wide differences in internet provisioning, packet jitter and losses due to network congestion or excessively delayed transmissions are generally unavoidable and require some form of mitigation. PLC methods for real-time multimedia streaming have been widely studied \cite{7870757} and recently ML approaches for speech reconstruction using neural networks have been proposed \cite{lee2015packet,lotfidereshgi2018speech}. 

Of the studies focused on NMP applications, the majority treat error recovery for transmission of MIDI signals. The authors of \cite{lazzaro2001case} propose a PLC approach that relies on the RTP protocol and its associated transport control protocol RTCP. The solution leverages a recovery journal section within the RTP packet which holds information a receiver uses to recover from earlier lost packets. This way, the ﬁrst packet received after one or multiple consecutive packet losses enables the recovery from all artifacts caused by missing audio data. The drawback of this approach is that it generates transmission overheads. Other recovery approaches for MIDI signals rely on auxiliary reliably-connected channels to transmit critical MIDI events \cite{virolainen2005methods} or implement acknowledgment mechanisms to allow for re-transmission of lost packets \cite{nelson2017reliable}.

In \cite{wang2008system}, a method for the detection of beat patterns of music signals is discussed as a means for prediction. At the sender side, the beat information is included as ancillary data to a preceding audio data interval in the transmitted compressed audio stream and is leveraged to perform PLC at the receiver. The method incurs transmission overhead.
The authors of \cite{fink2014low} propose a low-cost period extraction and alignment module to synthesize concealment signals from previous raw audio data, thus avoiding the use of autoregressive models in order to reduce the computation time. Period extraction is based on zero-crossings and matched pre-processing. To ensure smooth transitions, the extrapolated blocks are cross-faded from/into the previous/following audio frames. 
The main difference between the method proposed in this paper and the one described in \cite{fink2014low} is that, instead of repeating periods extracted from the previously received signal, we synthesize the waveform to be inserted in place of a missing packet. There has also been prior work which incorporates time scale modification techniques for audio concealment \cite{sanneck1996new}. The authors of \cite{sanneck1996new} compared their method with existing techniques (silence substitution and pattern replication). A next wave of PLC techniques based upon AR models were shown to outperform these approaches \cite{zhang2008autoregressive}. These methods are similar to infill approaches in the image domain \cite{oord2016pixel} which try to predict raw values of missing pixels based upon the neighboring content of an image.

The approach described here looks at both the spectrogram and the waveform of the input to predict the upcoming time-domain signal within a short interval. The method makes use of a latent space representation which summarizes the content of the signal in order to predict the next audio packet. The advantage of using latent space-based approaches is that a suitable summary representation can be learned according to the problem of interest, e.g., to simultaneously reconstruct a speech signal and to decode its transcription \cite{haque2019audio} or to transform audio signals from a given domain to another domain \cite{haque2018conditional}. Additionally, latent representations have been used to reconstruct the signal of interest in problems such as style transfer \cite{verma2018neural} and voice transformation \cite{van2017neural} by successfully mapping contents and characteristics of audio representations onto lower-dimensional vectors. 

%The advantage of using a latent space based approach is that it can be learned according to the problem of interest, which in our case summarization of the contents of the signal to predict the future packet. \cite{haque2019audio} learned summarization of the audio signal to simultaneously reconstruct back the signal, and decode back the contents of the audio signal i.e.\ the spoken letters. \cite{haque2018conditional} used latent summaries to transform audio signal from one domain of interest to another domain.

\section{Background on Autoregressive Prediction Models} \label{sec:background}

AutoRegressive (AR) models of stochastic processes are described by a set of parameters which are the coefficients of the linear regression of the current output random variable against its previous values. By assuming such a model for the transmitted audio signal, an AR model of order $p$, denoted by AR($p$), can be effectively used to forecast the future behavior of the time series as a linear combination of its past $p$ samples.
The regression parameters can be computed and dynamically updated using the most recently received packets (usually the last valid buffer) by, for example, solving an Ordinary Least Square (OLS) problem or as the solution of the well-known Yule-Walker equations. 
Since speech signals are often quasi-stationary within short-time analysis windows, AR models have been found to be useful in predicting the content of missing packets \cite{kondo2006speech,zhang2008autoregressive}. Similarly, music signals are generally well behaved and analogous assumptions can be made, especially in the case of recordings of a single pitched instrument. Therefore, we use this simple approach as a baseline to evaluate the performance of our ML model. Notice, however, that various extensions to the classical AR model have been proposed in the literature, such as the inclusion of a long-term predictor whose lag is tuned according to the estimated pitch. For a detailed overview of AR models and other signal processing approaches, readers are advised to refer to \cite{perkins1998survey}.

\section{Deep Learning Framework for Low-Latency Audio Prediction} \label{sec:framework}

\begin{figure*}[tbh]
\begin{minipage}[t]{1.0\linewidth}
  \centering
  \centerline{\includegraphics[width=0.85\linewidth]{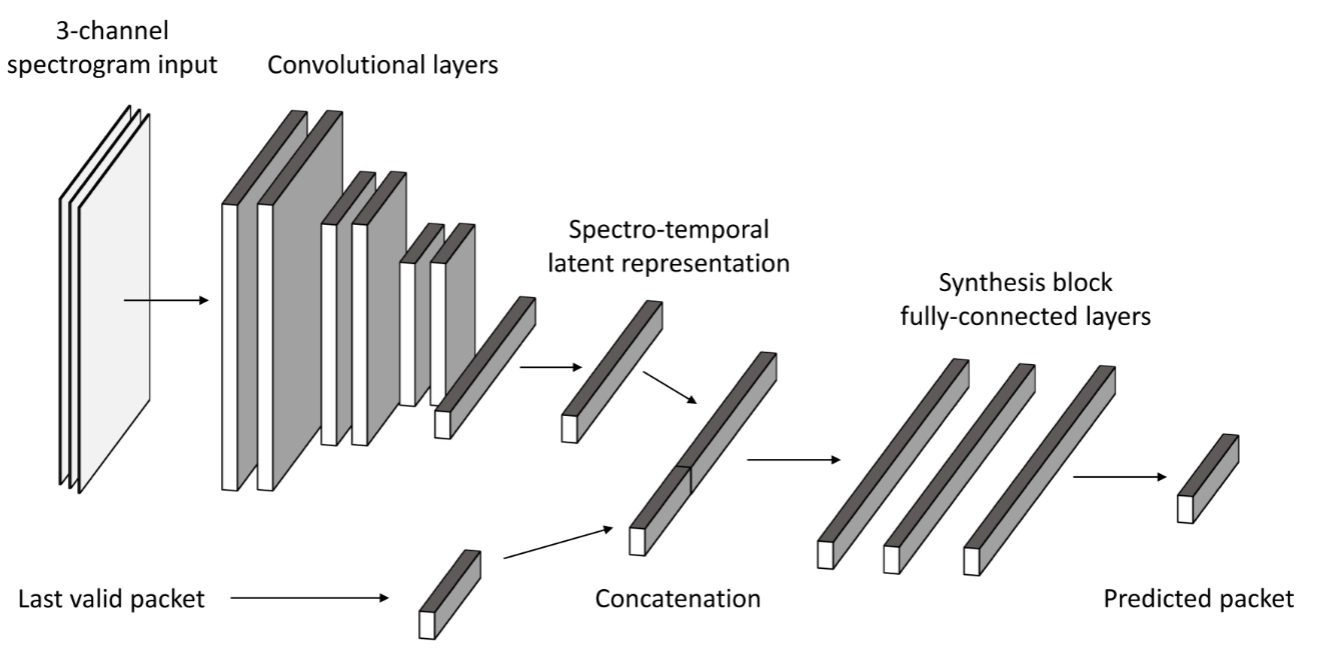}}
%  \vspace{2.0cm}
\end{minipage}
\caption{Architecture of the proposed method for synthesizing new audio packets using a neural network.
}
\label{fig:architecture}
%\vspace{0.75em}
%
\end{figure*}

Modeling longer term dependencies is considered a hard ML problem, especially for time-domain audio signals \cite{dieleman2018challenge}. In the literature, a similar problem has been addressed in the field of computer vision, when trying to reconstruct a missing part of an image \cite{oord2016pixel}. The high sample rates of digital audio signals present a significant challenge for learning small compact representations of long sequences, even for durations of 2--3 seconds. 

We adopt an hybrid approach which makes the problem tractable by combining both spectral and time-domain signal representations. The hypothesis of our approach is as follows: for a missing packet, the correctly predicted waveform will depend on what has happened in the past as well as the smoothness of the reconstructed signal. The latter requirement refers to the continuity of the signal derivatives at the edges between subsequent packets which need to be smooth in order to avoid audible clicks or glitches. As pointed out in \cite{dieleman2018challenge}, it is difficult to model waveforms on the scale of a few seconds, as these can be time sequences on the order of 50k samples or greater (considering 16~kHz as the sampling rate). In order to learn the behavior of time-domain signals characterized by longer-term dependencies and predict packets to come, we utilize spectro-temporal representations of waveforms in a training database. The same transform is applied to already received packets which are stored in a first-in first-out buffer. We compute a 100-bin mel-spectrogram of the previous 2~s of audio using a hop size of 10~ms and a 30~ms long Hanning window. A representation frame of 100$\times$200 corresponds to the past 2~s of audio. We also extended the previous context to 4~s and 8~s and  sub-sample the mel-spectrograms by a factor of 2 and 4, respectively, to yield the same representation frame size. All three frames (2~s, 4~s and 8~s) are stacked into 3 channels of a real-valued tensor of size 100$\times$200$\times$3. The mel-spectrogram is magnitude only (which throws away the phase/relative shift of the time-domain signal). Our assumption is that only the last valid packet matters when we synthesize the missing future packet in order to 
ensure the smoothness of the reconstructed waveform.
% maintain the continuity of the time-domain signal.

The first part of our hybrid approach uses a convolutional neural network which takes as input the 100$\times$200$\times$3 spectro-temporal representation and produces a 512-dimensional vector. This greater than 100-fold dimensional reduction proceeds as follows: a 3-layer convolutional architecture with 3$\times$3 kernels and 1$\times$2 pooling is connected to 2$\times$2 pooling in the next two layers and finally a linear layer reduces the dimensionality to 512. Each of the convolutional layers has 64 channels. 

The second part of our hybrid approach synthesizes a valid packet, conditioned on the latent code as extracted from the past signal history by the convolutional neural network. It concatenates the immediately previous valid packet of size 128 with the latent code of the past spectro-temporal content of size 512. These two signals, one for learning continuity in the time domain and one for the spectral content, are concatenated to produce a summary vector which contains both aspects of the signal. This is fed to a synthesis neural network, consisting of 3 fully-connected layers of 1024 neurons, and which is used to predict the missing or lost packet as depicted in Fig.~\ref{fig:architecture}.
%%%% [Alessandro] I believe that this shouldn't go here, but rather in the conclusions/future work! %%%
%
%In future, this can be replaced by a state of the art RNN module, that can be conditioned on the latent summary vector.

We minimize the $L_1$ norm of the difference between the predicted packet and the target one. The latent summary vector, encapsulating the spectral content of the past, along with the synthesis block that generates the missing packet, are all learn-able from the data by adjusting the network weights via backpropagation. The Adam optimizer  \cite{kingma2014adam} was used for the network training with learning rate decay starting from $10^{-3}$ and then lowered to $10^{-7}$ over the course of 100 epochs . The $L_1$ norm was seen to have better convergence than the $L_2$ norm, one of the reasons being that all of the signals are in the range of $[-1,1]$.

Not every audio sample point is equal with regards to error when stitching the predicted packet into an audio stream. In terms of continuity, errors at the beginning and at the end of a packet are much more perceptually significant than those in the center. Those at the edges introduce discontinuities in the waveform and its derivatives which result in audible artifacts. As one possible solution, we include a weighted loss function which penalizes the prediction error of those samples that are found on the edges of a packet more than of those in the center. This is achieved by multiplying the absolute error for each sample by an inverse Hanning window:
\begin{equation*}
    w[n] = 1 - \sin^2 \left( \frac{\pi n}{N-1} \right)
    \label{eq:inverseHann}
\end{equation*}
where $n=0,1,2,...,N-1$ is the sample index within a packet and $N=128$ is the packet length.
This approach is very similar to other nonlinear weighting techniques, for example, C-loss, Hinge Loss and other approaches used to handle data imbalances in machine learning \cite{wu2007robust}. Our assumption is that the network, by focusing on the edge regions of each packet, will autonomously learn how to generate seamless transitions.

\section{Performance Assessment} \label{sec:results}
\subsection{Database and Audio Acquisition Setup}
For the purpose of this paper, all of the results are reported for a single instrument, violoncello. The recordings were made in a consistent manner with the same player and music and the same acoustical setup (microphone, positions and room). This was done to reduce the influence of these factors on our modeling results.
A total of 15 hours of audio recording was collected. The cellist was practicing Bach's Six Suites for Violoncello over the course of a few weeks. These are extremely raw, unedited recordings including instrument tuning, scales practice and occasional improvisations. For the sake of the experiment, the database can be regarded as representative of a rehearsal of classical music such as might take place in a NMP session. The room was fairly reverberant with a high ceiling and sparsely furnished. The dataset is available online \cite{celloData}.

%%%% [Alessandro] I believe that this shouldn't go here, but rather in the conclusions/future work! %%%
%
%In the future, a diverse dataset can be collected that will reflective of the kind of performance that can be expected in an real-world setting, or adapted from availability of large scale unlabelled datasets such as YouTube. We would expect the performance of machine learning system to drop, when other instrument which are not present in the training set are used for testing.

For the autoregressive approach, since it only requires storing of a set of past buffers, we do not create training data. For neural networks, we need to build a training and validation set to optimize the parameters. 

Using a high-speed wired LAN with a gateway router coupled to WAN modem for either fiber-to-the-home, digital subscriber line or cable access to the Internet, packet dropouts usually happen only as a very small percentage of the total number of packets transmitted. Statistical distributions of packet loss probabilities are extremely variable, as they are highly depending on network traffic conditions. 
%This makes it difficult to collect data from a live situation as it is a stochastic event and would require enormously long captures. 
Therefore, for training purposes, we presume that packet dropouts are isolated and can happen at any instant. We sample time windows corresponding to the content carried by one packet at random points in the cello recordings. 
In particular, the missing packets thus selected amount to 1\% of the available audio data. For each packet, we extract the respective spectro-temporal context and store the previous valid packet.
A point to note is that a higher sampling would yield more training data and thus improved performance. This choice, however, was mainly due to the limited computational resources at our disposal.

The experiments use contexts of either 2~s or 8~s (storing intermediate representations of 2~s, 4~s and 8~s into 3 channels) that were (down-)sampled at 16~kHz (from 48~kHz) and a packet size of 128 samples. The goal of the experiments is: given the previous valid packets which have been received, can we predict the next packet?

Data input from the training and test sets were normalized to have a unity peak amplitude at each hop. For example, in the 2~s context case each new 2~s window-full would be normalized.
%
% We sample 1\% of the data as training, i.e., spectro-temporal context, previous packet and trying to predict the future packet.
%A point to note is that higher portion of sampling would yield improved performance, as seen across ML problems. This choice was mainly to keep iterating faster over the training data available with us, and the computation resources at our disposal.
%
%%% It is difficult to mimick the missing packets, hence we presumed that packet dropout can happen at any instance for the collection of the dataset. 
By minimizing the $L_1$ norm between the predicted and the target waveforms, the model is biased towards reducing the errors of higher amplitude signals rather than lower amplitude signals.
%The same error $L_1$ norm error, can be significant for higher amplitude signals than with lower amplitude packets without normalization.
Context normalization avoids this amplitude-dependent performance bias.

Unfortunately, this also boosts packets which only have low-amplitude background noise as their content. However, since only a small fraction of packets in the database are noise-like, the neural network will mainly learn how to model deterministic signals, which in this case are mostly music signals, and will apply the learned model to noise-like signals, too. 

\subsection{Numerical Evaluation}
We evaluate the latent space-based ML method described above and compare it to a baseline AR($p$) model implemented using the \textit{statsmodels} Python library \cite{seabold2010statsmodels}. Only isolated, single packet dropouts have been tested to date. Evaluation was performed on a held out test set. %containing different recordings than the training set.
This test data consists of a single track of just over 11 minutes that was not present in the training dataset, for a total of 646 missing packets.
%, which consisted of a single track not present in our dataset for evaluation.
As mentioned earlier, this test set vs.\ training set split does not affect AR($p$) performance, as AR models estimate coefficients only from the previous history of the signal. An advantage of the AR model is that it can update its parameters from observing short-time correlations in the input stream, whereas ML methods derive a fixed set of parameters from \textit{all} of the entire training set.

In comparing algorithm performance using context time scales of 2~s vs.\ 8~s, we noted that longer contexts do not improve the validation loss.
%Consequently, 2~s was used for all further experiments. 
For the comparison, we trained the network shown in Fig.~\ref{fig:architecture} using two different kinds of spectro-temporal inputs. The first one consisted of the 2~s, 4~s and 8~s representation frames stacked in the first, second and third input channel of the convolutional neural network, respectively. The second, for compatibility with the same model architecture, consisted of the 2~s mel-spectrogram replicated it in all three input channels.
The ML method provided equivalent results in both test cases. Therefore, the shorter 2~s context was used for all further experiments. 
%There can be two variants of Fig.~\ref{fig:architecture}, one with the three channels being having context of 2~s, 4~s and 8~s and the other version just having a context of 2~s. For compatibility with the model architecture as shown in  Fig.~\ref{fig:architecture}, we tiled the 2~s mel-spectrogram representation by replicating it in all the three channels.
This finding indicates that PLC is not a long-term dependency problem, a fact reinforced also by the competence of the AR model in our testing scenario. 

All of the results were evaluated on a Tesla K80 GPU using TensorFlow 1.15. In terms of achieving an eventual real-time system, things look favorable. The GPU-based system ran roughly 10 times faster than real time when computing the mel-spectrogram and doing a forward pass using the trained system. With a CPU-based system, the forward pass could not be computed in real time (defined as completing the process within the interval between the arrival of two packets). In our case, with a 1.8~GHz Intel Core i5 processor, it ran approximately 5 times slower than real time. Future improvements in both computational power and in the architectural simplification of deep learning-based models make it hopeful that such models can become tractable in real time without relying on hardware acceleration.

While perceptually-informed objective measures such as PEAQ \cite{thiede2000peaq} have been developed for assessing the quality of audio codecs, there is yet no consensus among the research community on whether these methods could prove reliable in evaluating the effectiveness of PLC algorithms \cite{fink2013comparison, khalifeh2017}. 
%Algorithm performance is hence measured as the average absolute error between the predicted waveforms and actual ones. 
%
Hence, we measure the performance of both the proposed ML method and the AR baseline as the average absolute error between the predicted waveforms and actual ones.
As described above, all tests involve normalized waveforms such that the peak is always one. This criterion avoids dependency on the amplitude of the signal. In the testing scenario described, we see that the ML-based approach outperforms the AR model within certain regimes. 

\begin{figure}[t]
\begin{minipage}[b]{1.0\linewidth}
  \centering
  \centerline{\includegraphics[width=1.0\linewidth]{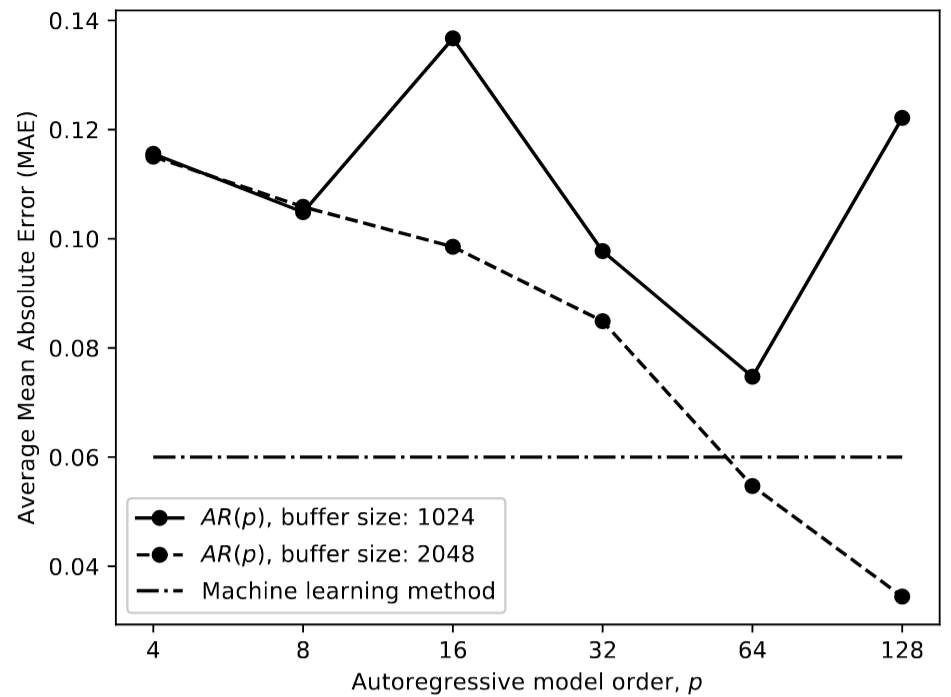}}
%  \vspace{2.0cm}
\end{minipage}
\caption{Average Mean Absolute Error (MAE) of the autoregressive model AR($p$) as a function of the number $p$ of regression coefficients estimated using the last 1024 and 2048 valid samples (solid and dashed line, respectively) plotted against the average MAE of the proposed machine learning method (dash-dotted line).}
\label{fig:mae}
\end{figure}

The 128 sample packet size used in the experiment represents 8~ms of audio. The AR model comparison was run with 1024 and 2048 samples of history (64~ms and 128~ms, respectively) which were used for the conditional maximum likelihood estimation of the regression parameters via OLS. Its performance therefore depends on both the amount of history and the order $p$ of the filter applied. AR($p$) only achieves parity with the ML method with a 128~ms window and order above 57 as seen in Fig.~\ref{fig:mae}. In our experiments, the AR model appears effective in predicting short gaps of missing audio (128 samples). However, AR($p$) is able to outperform the ML method only with a fairly high number of regression parameters. If we were to test the PLC system on the longer gaps of bursty packet losses, then either the number of parameters to be estimated and thus the amount of history required would quickly become unmanageable or the results may converge to a zero signal after a few ms (a longer-term tendency for AR models). In this respect, ML-based approaches could provide an alternative solution.

\section{Conclusion}
\label{sec:conclusion}

We described a neural network-based technique for packet loss concealment, specifically designed to be implemented in real-time audio streaming applications for networked music performance.
The novel hybrid approach proposed in this paper marries two key elements. The first one is the spectral carry-forward of short-term (2~s) mel-spectrum content from a sufficiently large database of music recordings using latent vectors. The second one is a technique for the correct synthesis of a missing packet conditioned on a latent code. Used together, these methods predict and generate a packet which can be inserted in place of missing audio data. We believe that with future improvements to computational power, machine learning-based PLC will become pervasive for all types of audio signals. Investigations of possible improvements to the latent representation, as well as possible modifications of the synthesis block are envisioned, possibly by replacing the fully-connected neural network with a state-of-the-art recurrent module. Validating the experimental results via subjective listening tests in a real-world NMP scenario, generalizing the approach to tackle a wide variety of audio signals (including different musical instruments, genres, players and recording environments) and extending it to prediction of longer excerpts (for example, those occurring in the case of bursty packet losses) will also be objectives of further studies. 
%We can also evaluate the current system in real world setup for future evaluations. 

\section*{Acknowledgements}
We acknowledge the advice and encouragement of Andrea Bianco and thank Klaus Greff for the useful discussions. 

% Below is an example of how to insert images. Delete the ``\vspace'' line,
% uncomment the preceding line ``\centerline...'' and replace ``imageX.ps''
% with a suitable PostScript file name.
% -------------------------------------------------------------------------

% To start a new column (but not a new page) and help balance the last-page
% column length use \vfill\pagebreak.
% -------------------------------------------------------------------------
%\vfill\pagebreak

% References should be produced using the bibtex program from suitable
% BiBTeX files (here: strings, refs, manuals). The IEEEbib.bst bibliography
% style file from IEEE produces unsorted bibliography list.
% -------------------------------------------------------------------------
\bibliographystyle{IEEEbib}
\bibliography{strings,refs}

\end{document}